\documentclass[dvips,12pt]{article}
\usepackage{graphicx}
\usepackage{repko}
\usepackage{epsf}
\def\ds{\displaystyle}

\def\sb{\mbox{\rule{0pt}{11pt}}}

\def\Dot{\!\cdot\!}
\def\al{\alpha}

\def\de{\delta}
\def\la{\lambda}
\def\ro{\rho}
\def\ep{\varepsilon}

\newfont{\ex}{cmr10}
\def\hb{\mbox{\rule{16pt}{0pt}}}

\def\t#1{$\times 10^{#1}$}
\def\sla#1{#1\hspace*{-5pt}/}
\def\pa{\partial}
\begin{document}

\title{\large\bf Plasma Energy Loss into Kaluza-Klein Modes}
\author{\normalsize Duane A. Dicus$^1$, Wayne W. Repko$^2$, and Vigdor L. Teplitz$^{3,4}$\\
{\small\it $^1$Center for Particle Physics, University of Texas, Austin, TX
78712}\\{\small\it$^2$Department of Physics and Astronomy, Michigan State
University, East Lansing, MI 48824}\\{\small\it $^3$Department of Physics,
Southern Methodist University, Dallas, TX 75275.}\\{\small\it $^4$Office of
Science and Technology Policy, Executive Office of the President, Washington,
DC 20001\footnote{Address until June 30, 2001}}}

\date{\normalsize\today}
\maketitle
\begin{abstract}
Recently, Barger {\em et al.} computed energy losses into Kaluza Klein modes
from astrophysical plasmas in the approximation of zero density for the
plasmas. We extend their work by considering the effects of finite density for
two plasmon processes. Our results show that, for fixed temperature, the energy
loss rate per cm$^3$ is constant up to some critical density and then falls
exponentially. This is true for transverse and longitudinal plasmons in both
the direct and crossed channels over a wide range of temperature and density. A
difficulty in deriving the appropriate covariant interaction energy at finite
density and temperature is addressed. We find that, for the cases considered by
Barger {\em et al.}, the zero density approximation and the neglect of other
plasmon processes is justified to better than an order of magnitude.
\end{abstract}

\section{Introduction}

    Recently, Barger {\em et al.} \cite{betal} addressed astophysical
constraints on extra dimensions by computing energy loss rates from the sun,
red giant stars and (Type II) supernovae due to possible excitation of graviton
modes, $\mathcal{G}$, in the case that the extra dimensions are compactified
\cite{add}. The processes $\gamma \gamma\to \mathcal{G},\, e^+e^-\to
\mathcal{G}, \gamma e\to\mathcal{G},\, e N\to\mathcal{G} e N$ (in the static
nucleon approximation), and $NN\to\mathcal{G}NN$ where considered. They worked
in the zero density approximation, varying only the temperature. Their
calculation neglected plasma effects and they anticipated that this neglect
should not be important because of the high power dependence on $M_S$, the
inverse of the compactification dimension.

    The purpose of this paper is to address the extent to which the
process $\gamma \gamma\to\mathcal{G}$ (and the crossed process
$\gamma\to\gamma\,\mathcal{G}$) is affected by a non-zero charged particle
density and the presence of both longitudinal and transverse plasmons.  Our
aims are, in brief, to find numerical values for the energy loss rate per
cm$^3$ for densities from $1$ to $10^{15}$\,gm/cm$^3$ and temperatures from $1$
to $10^9$\, eV, to confirm the Barger {\em et al.} expectation of density
insensitivity, to show the extent to which the systems they considered are
close to the border in density at which their expectation fails, to determine
the relative contributions of longitudinal and transverse plasmon processes, to
address the size of the contribution of the crossed process, and, importantly,
to note the ambiguities in the form of a covariant interaction between plasmons
and Kaluza-Klein modes.

    It is clear that the expectation of Barger et al. \cite{betal}
must fail at sufficiently high density, for fixed temperature, since the energy
loss rate goes as an integral over the Bose-Einstein distribution, $(e^{\omega
_i/kT}-1)^{-1}$, in which the frequencies are given by a dispersion relation
with an effective photon mass that grows with density. It is essentially a
numerical question as to the point at which suppression sets in and the rate
(in density) at which it proceeds. Intuition is hampered by the fact that the
natural parameter is the electron chemical potential which is not simply
related to the particle density.  We give the numerical results over the
temperature and density ranges cited above.

    An important problem in carrying out this work is the question of
the appropriate Lagrangian.  The free space coupling between the
electromagnetic field and gravitons can be found in textbooks, for example
\cite{weinberg}, and has been generalized to the case of higher dimensional
Kaluza-Klein excitations \cite{grw,hlz}.  However, we have not found a parallel
literature for the case in which the free space photon is replaced by a plasmon
satisfying a non-trivial dispersion relation. This difficulty is addressed in
Section 2. We adopt a diagrammatic approach and also make an approximation that
we test numerically.  Also given in Section 2 is the formalism used for the
numerical calculations of Section 3.  We conclude in Section 4 with a brief
summary.

\section{Formalism}

In a medium with nonzero temperature and density, radiation satisfies the
dispersion relation
\begin{equation}
\omega^2-|\vec{k}|^2 = \Pi_a(\omega,|\vec{k}|)\,.
\end{equation}
$\Pi_a(\omega,|\vec{k}|)$ is the transverse or longitudinal component of the
polarization tensor,
\begin{equation}
\Pi_a(\omega,|\vec{k}|)=\ep^{\mu\;*}_a\Pi_{\mu\nu}\ep^{\nu}_a\,,
\end{equation}
where $\ep^{\mu}_a$ ($a=L,T$) are the polarization vectors and the polarization
tensor $\Pi_{\mu\nu}$ is the photon self energy in the medium. The contribution
to this self energy from fermions in the medium is calculated in the medium
rest frame by adding a term
\begin{equation}\label{prop}
\frac{\sla{p}+m}{e^{(E\pm\mu)/T}+1}2\pi i\de(p^2-m^2)
\end{equation}
to the usual (vacuum) progagator. Here, $T$ is the temperature, $E=p^0$ and
$\mu$ is the electron chemical potential, which is related to the electron
number density $n_e$ by
\begin{equation}\label{density}
2\int\frac{d^{\,3}p}{(2\pi)^3}\left[\frac{1}{e^{(E-\mu)/T}+1} -
\frac{1}{e^{(E+\mu)/T}+1}\right]= n_e
\end{equation}
To lowest order in the fine structure constant $\al$, the polarization tensor
is given by
\begin{eqnarray}\label{pimunu}
\Pi^{\mu\nu} &=&
16\pi\al\int\frac{d^{\,3}p}{(2\pi)^32E}\left[\frac{1}{e^{(E-\mu)/T}+1} +
\frac{1}{e^{(E+\mu)/T}+1}\right]\nonumber \\[4pt] & &\qquad\times\frac{(p\Dot
k)^2g^{\mu\nu}+k^2p^{\mu}p^{\nu}-p\Dot k(k^{\mu}p^{\nu}+
k^{\nu}p^{\mu})}{(p\Dot k)^2}\,.
\end{eqnarray}

It turns out that $\Pi_{L}$ and $\Pi_{T}$ can be approximated to within 1\% for
all temperatures and densities \cite{bs} by
\begin{eqnarray}
\Pi_L &=&
\omega_P^2\left[1-G(v_*^2|\vec{k}|^2/\omega^2)\right]+v_*^2|\vec{k}|^2-|\vec{k}|^2\,,
\\ [4pt]
\Pi_T &=&\omega_P^2\left[1+\frac{1}{2}G(v_*^2|\vec{k}|^2/\omega^2)\right]\,,
\end{eqnarray}
where $v_*$ is an average value of $v=|\vec{p}|/E$ for the electron (the only
fermion which contributes for stellar temperatures and densities). Explicitly,
\begin{equation}
v_*=\omega_1/\omega_P\,,
\end{equation}
with $\omega_1$ given by
\begin{equation}
\omega_1^2=\frac{4\al}{\pi}\int_0^{\infty}\,d|\vec{p}||\vec{p}|\left(
\frac{5}{3}v^3-v^5\right)f_{E}\,,
\end{equation}
where $f_E$ is the sum of the electron and positron distributions (the square
bracket in Eq.\,(\ref{pimunu}) above). The plasma frequency $\omega_P$ is given
by
\begin{equation}
\omega_P^2=\frac{4\al}{\pi}\int_0^{\infty}\,d|\vec{p}||\vec{p}|\left(
v-\frac{1}{3}v^3\right)f_{E}\,,
\end{equation}
and the function $G(x)$ is
\begin{equation}
G(x)=\frac{3}{x}\left[1-\frac{2x}{3}-\frac{1-x}{2\sqrt{x}}\log\left(\frac{1+\ds\sqrt{x}}
{1-\ds\sqrt{x}}\right)\,\right]\,.
\end{equation}

It will be important below to note that, for transverse photons, $k^2 =
\omega^2-|\vec{k}|^2$ is $\omega_P^2$ for $|\vec{k}|=0$ and {\em increases} as
$|\vec{k}|$ increases, while, for longitudinal photons, $k^2=\omega_P^2$ at
$|\vec{k}|=0$ and {\em decreases} as $|\vec{k}|$ increases. Integration over
$|\vec{k}|$ for longitudinal photons must be cut off at the point where $k^2$
becomes negative,
\begin{equation}\label{kmax}
|\vec{k}|^2_{\rm
max}=\frac{3\omega_P^2}{v_*^2}\left[\frac{1}{2v_*^2}\log\left(\frac{1+v_*}{1-v_*}
\right)\right]\,.
\end{equation}

We include in our numerical evaluations the renormalization constants
\begin{equation}
Z^{-1}_a = 1-\frac{\pa\Pi_a}{\pa\omega^2}\,,
\end{equation}
although this is inconsistent with calculating only to the lowest order in
$\al$. It is a check on our results that they do not change significantly when
the $Z_a$ are set to unity. The $Z_a$ are given by \cite{raffelt}
\begin{eqnarray}
Z_T & = &
\frac{2\omega^2(\omega^2-v_*^2|\vec{k}|^2)}{\omega^2[2\omega_P^2-2(\omega^2-|\vec{k}|^2)]
+(\omega^2+|\vec{k}|^2)(\omega^2-v_*^2|\vec{k}|^2)} \\ [4pt]
Z_L & =
&\frac{2\omega^2(\omega^2-v_*^2|\vec{k}|^2)}{3\omega_P^2-\omega^2+v_*^2|\vec{k}|^2}
\frac{\omega^2}{\omega^2-|\vec{k}|^2}\,.
\end{eqnarray}

The rate of graviton emission can be calculated using the Lagrangian for the
coupling of Kaluza-Klein field ${\cal G}_{\vec{n}}^{\mu\nu}$, corresponding to
the mass excitation $m_{\vec{n}}^2=(2\pi)^2\vec{n}^2/R^2$, to the photon
energy-momentum tensor $T_{\mu\nu}$. Neglecting gauge terms, this coupling is
\cite{grw,hlz}
\begin{eqnarray}\label{lint}
{\cal L} & = &-\frac{\kappa}{2}{\cal G}_{\vec{n}}^{\mu\nu}T_{\mu\nu} \nonumber \\ [4pt]
& = &\frac{\kappa}{2}\left({\cal G}_{\vec{n}}^{\mu\nu}F_{\mu}^{\la}F_{\nu\la}-\frac{1}{4}
{\cal G}_{\vec{n},\mu}^{\mu}F^{\la\ro}F_{\la\ro}\right)\,,
\end{eqnarray}
where $F_{\mu\nu}$ is the electromagnetic field tensor. We consider only the
coupling of the spin-2 component of the Kaluza-Klein field; the spin-0
component does not couple to photons.

The matrix element for $\gamma(k_1)\gamma(k_2)\to{\cal G}$ obtained from
Eq.\,(\ref{lint}) is \cite{grw,hlz}
\begin{equation}\label{matrixel}
{\cal M} =
\frac{\kappa}{2}\ep^{\la}(k_1)\ep^{\ro}(k_2)\ep^{*\,\mu\nu}(k_1+k_2)T^{(0)}_{\mu\nu,\la\ro}\,,
\end{equation}
where
\begin{equation}\label{coupling}
T^{(0)}_{\mu\nu,\la\ro}=k_1\Dot
k_2C_{\mu\nu,\la\ro}+D_{\mu\nu,\la\ro}(k_1,k_2)\,,
\end{equation}
with
\begin{eqnarray}
C_{\mu\nu,\la\ro}& = &
\eta_{\mu\la}\eta_{\nu\ro}+\eta_{\mu\ro}\eta_{\nu\la}-\eta_{\mu\nu}\eta_{\la\ro}
\\ [4pt]
D_{\mu\nu,\la\ro}(k_1,k_2)& = & \eta_{\mu\nu}k_{1\,\ro}k_{2\,\la} -
\left[\eta_{\mu\ro}k_{1\,\nu}k_{2\,\la}+\eta_{\mu\la}k_{1\,\la}k_{2\,\nu}
+ (\mu\leftrightarrow\nu)\right]\,.
\end{eqnarray}

The sum over polarizations of the Kaluza-Klein state is \cite{grw,hlz}
\begin{equation}\label{spinsum}
\sum_{s=1}^5\,\ep_{\mu\nu}^{s}(k)\ep_{\la\ro}^{s\,*}(k)=\frac{1}{2}B_{\mu\nu,\la\ro}(k)\,,
\end{equation}
with
\begin{eqnarray}
B_{\mu\nu,\la\ro}(k)& =
&E_{\mu\la}E_{\nu\ro}+E_{\mu\ro}E_{\nu\la}-\frac{2}{3}E_{\mu\nu}E_{\la\ro} \\
[4pt] E_{\mu\nu}& = &
\eta_{\mu\nu}-\frac{k_{\mu}k_{\nu}}{m_{\vec{n}}^2}\,.\label{projop}
\end{eqnarray}

The coupling Eq.\,(\ref{coupling}) is gauge invariant even if $k_1^2$ and
$k_2^2$ are not zero, e.g. $k_1^{\mu}T^{(0)}_{\mu\nu,\la\ro}=0$. However, it is
not conserved, $(k_1+k_2)^{\mu}T^{(0)}_{\mu\nu,\la\ro}\neq 0$, if $k_1^2$
and/or $k_2^2$ differs from zero. We cannot write a conserved coupling by using
the energy-momentum tensor for a massive vector field because $k_1^2$ is not
necessarily equal to $k_2^2$. This means that if we square ${\cal M}$ of
Eq.\,(\ref{matrixel}), and use Eq.\,(\ref{spinsum}), we get extra terms of the
form $k_1^2/m_{\vec{n}}^2$ or $k_2^2/m_{\vec{n}}^2$ from the second term in
Eq.\,(\ref{projop}).

\begin{figure}[h]
\centering\includegraphics[height=3.0in,clip]{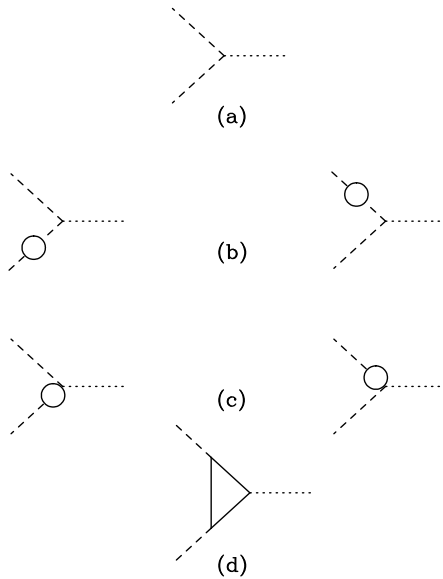}
\caption{\footnotesize The Feynman diagrams for the one-loop corrections to the
energy-momentum tensor are shown. The dotted lines denote Kaluza-Klein
gravitons, the dashed lines photons and the solid lines electrons.
\label{diags}}
\end{figure}

To have a conserved amplitude with $k_1^2\neq k_2^2\neq 0$, we must include all
the diagrams of Fig.\,(\ref{diags}). The Feynman rules for the
fermion-fermion-${\cal G}$ coupling and the fermion-fermion-photon-${\cal G}$
are given in Refs.\,\cite{grw,hlz}, and the loops are calculated by using
Eq.\,(\ref{prop}) for one of the legs. We have shown the the sum of these
diagrams is gauge invariant and conserved for arbitrary $k_1^2$ and $k_2^2$ at
finite temperature and density. However, this was done without actually
evaluating the diagrams. In particular, diagram (d) is very tedious and we have
not computed it. Instead, we have used only diagram (a) (Eq.\,(\ref{matrixel})
above) but have evaluated every energy loss twice - once including the
$k_1^2/m_{\vec{n}}^2$ and $k_2^2/m_{\vec{n}}^2$ terms and once omitting them.
In every case, the results were almost identical. While this proves nothing, it
does seem to indicate that performing the complete one-loop calculation would
not give a substantially different answer.

The reaction rate must be summed over the Kaluza-Klein states, which is done by
integrating over
\begin{equation}\label{dmsq}
dm_n^2\,\frac{4\pi m_n^{(n-2)}}{\kappa^2M_S^{n+2}}\,,
\end{equation}
where $n$ is the number of extra dimensions. $M_S$ is the string scale which is
related to the compactification  scale $R$ and Newton's constant $G_N$.
Specifically, we use
\begin{equation}
M_S^{n+2}R^n=\frac{(4\pi)^{n/2}\Gamma(n/2)}{4 G_N}\,.
\end{equation}
Our definition of $M_S^{n+2}$ differs from that of \cite{betal} by a factor of
$2$, i.e. their $M_S$ is larger by a factor $2^{1/(n+2)}$. As a consequence,
values of the energy loss per unit volume obtained from our tables must be
multiplied by 2 when comparing with Barger {\em et al.}\cite{betal}.

For 2 particles $\to$ 1 particle reactions there remains a delta function from
phase space which identifies $m_n^2$ with the center of mass squared energy
$s$. Thus, the integral over $m_n^2$, Eq.\,(\ref{dmsq}), replaces $m_n^2$ by
$s$ and our results depend on $n$ through the factor $s^{(n-2)/2}/M_S^{(n+2)}$.

The rate of energy loss per unit volume is given by the standard expression
\begin{equation}\label{dEdtdV}
{\cal Q}_{a,b}  =
A\int\frac{d^3k_1}{(2\pi)^3}\frac{1}{e^{\omega_1/T}-1}\int\frac{d^3k_2}{(2\pi)^3}
\frac{1}{e^{\omega_2/T}-1}(\omega_1+\omega_2)Z_a(|\vec{k}_1|)Z_b(|\vec{k}_2|)
v\sigma\,,
\end{equation}
where $v\sigma$ denotes the cross section times the relative velocity
\cite{betal}. The initial photons have $k_i^2 = \omega_i^2 - |\vec{k}_i|^2$, $i
= 1,2$ and can be transverse, $a=b=T$, longitudinal, $a=b=L$ or mixed, e.g.
$a=L,\,b=T$. The factor $A$ gives the number of spin states: $A=4,2,1$ for
$TT$, $TL$ or $LL$. For longitudinal photons, the $|\vec{k}|$ integrals are cut
off at $|\vec{k}|_{\rm max}$ given by Eq.\,(\ref{kmax}). The corresponding
expression for the energy loss in the decay $T\to L\,{\cal G}$ is
\begin{eqnarray}
{\cal Q}_{T\to L} & = & 2\int\frac{d^3k_T}{(2\pi)^32\omega_T}
\frac{1}{e^{\omega_T/T}-1}\int\frac{d^3k_L}{(2\pi)^32\omega_L}
\frac{(\omega_T-\omega_L)}{1-e^{-\omega_L/T}}Z_T(|\vec{k}_T|)Z_L(|\vec{k}_L|)
\nonumber \\ [4pt]
&  & \qquad\times\frac{(2\pi)^4\left((k_T-k_L)^2\right)^{(n-2)/2}}{\kappa^2M_S^{n+2}}
\,|{\cal M}|^2 \,,
\end{eqnarray}
where ${\cal M}$ is given by Eq.\,(\ref{matrixel}) with $k_2\to -k_L$.

\section{Calculations}

\begin{figure}[h]
\centering\includegraphics[height=3.0in,clip]{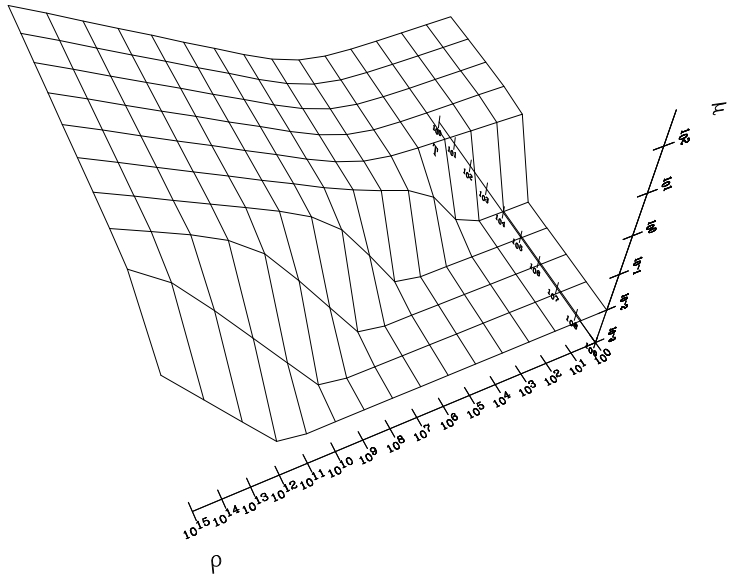}
\caption{\footnotesize
The chemical potential $\mu(\rho,T)$ is plotted for the range of density $\rho$
and temperature $T$ given in the text. Here, for display purposes, the lowest
value of $\mu$ shown, $10^{-2}$, is an upper limit on the exact numbers in
Table \ref{mu}.\label{mu_1}}
\end{figure}

The first step in the calculations is to obtain  $\mu (T,\rho)$ from
Eq.\,(\ref{density}).  In doing this we assume that the electron number
density, $n_e$ is related to the mass density $\rho$ by $n_e=\rho /m_p$ where
$m_p$ is the proton mass.  This is useful for comparison purposes and is a
reasonable order of magnitude approximation but needs correction (by less than
an order of magnitude) for a supernova or a neutron star.  The results of the
calculation of $\mu$ are given in Tables \ref{mu} and \ref{mutilde} for the
matrix of $\rho$ and $T$ values: $\rho =1.0,\,10.0,\,\ldots\, 10^{15}$\,
gm/cm$^3$ and $T=1,\,10^2,\,\ldots\, 10^9\; \mathrm{eV}$. Two tables are given
($\mu(T,\rho)$ and $\tilde{\mu}(T,\rho)=\mu(T,\rho)-m_e$) in order to make
clear both the deviation of $\mu$ from $m_e$ (taken as 0.51 MeV) at low
temperature and its deviation from zero at high.  Note the rapid variation of
$\mu$ (for the lowest densities) at the temperature $T$ around $0.1\;
\mathrm{MeV}$ where pair production first begins to be copious, and the slower
but similar variation at higher densities $\rho$. The variation is slower for
higher densities because the electron-positron density difference needs to have
a large value. These variations are illustrated in Fig.\,\ref{mu_1}, where, for
display purposes, the lowest value of $\mu$ shown, $10^{-2}$, is an upper limit
on the exact numbers in Table \ref{mu}.
In calculating $\mu$ we used an iteration procedure and required that the
output density value equal the input to better than a percent.

We now pass on to the results of calculating $\mathcal{Q}=d^4E/dtdV$ for the
case of $n=2$ extra dimensions. It was possible to evaluate the
integral in Eq.\,(\ref{dEdtdV}) over the cosine of the angle between the two
plasmons analytically so that, for all the processes under consideration, only
two integrals remain in finding the energy loss rate - the integrals over the
two plasmon momenta.

It should be noted that there are only four processes to consider: (1) $T+T\to
\mathcal{G}$, (2) $ T+L\to\mathcal{G}$, (3) $T\to L+\mathcal{G}$, and  (4)
$L+L\to\mathcal{G}$. This is because, as the plasmon momentum $|\vec{k}|$
increases, the effective mass of a transverse plasmon increases while the
effective mass of a longitudinal plasmon decreases \cite{raffelt}. Thus the
{\em missing} processes, $L\to T+\mathcal{G}$ and $L\to L+\mathcal{G}$ are
forbidden by energy-momentum conservation. The assertion is clear for the first
process since $m_T>m_L$. For the second, we note that, in the rest frame of the
decaying longitudinal plasmon, $m_L=\omega_P$ and conservation of energy and
momentum implies that the graviton mass, $m_{\cal G}$, satisfies
\begin{equation}\label{LneqLG}
m^2_{\cal G}=\left(\omega_P-\omega_L(|\vec{k}|)\right)^2-|\vec{k}|^2\,,
\end{equation}
where $\omega_L(|\vec{k}|)$ is the energy of the final plasmon. Using the
dispersion relation for longitudinal plasmons, it can be shown that the right
side of Eq.\,(\ref{LneqLG}) is less than zero for $|\vec{k}|>0$.

The results of the calculations are given in Tables \ref{TT}-\ref{Tot2}.  These
have the energy loss rates for the four processes, and for the sum, for a
matrix of density and temperature values - $1.0$ to $10^{15}$\, gm/cm$^3$ for
density $\rho$ and $1$ to $10^{9}$ eV for temperature $T$ - in both cases in
factors of $10$ increments.  In these tables, $T$ increases from left to right
while $\rho$ increases from top to bottom.  The entries are logs to the base 10
of the energy loss rate in ergs per cm$^3$-s.  Note that Barger {\em et al.}
\cite{betal} give results per unit mass, but results per unit volume are better
for our purposes since they show more clearly the way in which the zero-density
approximation breaks down as the density increases. We give the results for
$M_S=1$\, TeV. Two additional tables, \ref{dom} and \ref{frac}, give
respectively the number of the process that dominates for the reaction (zero if
the rate is zero, i.e. below $10^{-320}$) and the fraction of the total
represented by the dominant contribution.

\begin{figure}[h]
\centering\includegraphics[height=3.0in,clip]{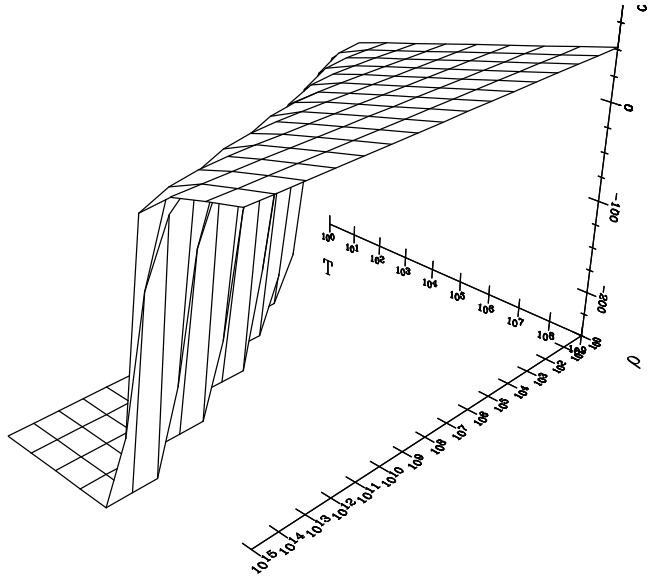}
\caption{\footnotesize The $\log_{10}(\mathcal{Q})$ for the process $T+T\protect\to\mathcal{G}$
is plotted for the range of density $\rho$ and temperature $T$ given in the
text. \label{tttog}}
\end{figure}

In Fig.\,\ref{tttog}\,($T+T\to\mathcal{G}$), we see at a glance the effect
cited in the
Introduction: for fixed $T$ the energy loss rate is independent of the density
until $\rho$ increases to a point where the effective photon mass and plasmon
density are sufficiently high that the rate drops exponentially. The numerical
values are given in Table\,\ref{TT}.
\begin{figure}[h]
\centering\includegraphics[height=3.0in,clip]{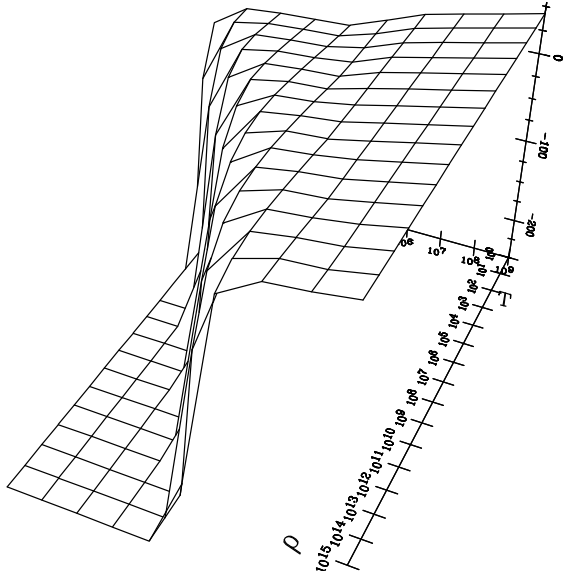}
\caption{\footnotesize The $\log_{10}(\mathcal{Q})$ for the process $L+L\protect\to\mathcal{G}$
is plotted for the range of density $\rho$ and temperature $T$ given in the
text. \label{lltog}}
\end{figure}
The analogous plot for $L+L\to\mathcal{G}$ is shown in Fig.\,\ref{lltog}. Here,
the the rate grows slightly for a fixed $T$ before dropping exponentially in
the mass of the longitudinal plasmon with increasing $|\vec{k}|$. Again,
numerical values are found in Table\,\ref{LL}.

The Sun, a red giant, and a Type II supernova are, in the ($\rho, T$) plane,
given by Barger {\em et al.} \cite{betal} to be at (156 gm/cm$^3$,\,1.3 keV),
(10$^6$ gm/cm$^3$,\, 8.6 keV), and (10$^{15}$ gm/cm$^3$,\, 30 MeV)
respectively. We see from Table \ref{TT} that the Sun is in a low density
region where the zero-density approximation holds while the supernova is on the
edge of a constant density region. We also see from Table 6 that the other
processes contribute little in these two cases. The red giant (RG) case is more
interesting.  In Table \ref{TT}, we see from the $T=10^4$ and $T=10^3$ columns
that the RG is in the gentle fall off region for the former, but the steep fall
off region for the latter. Examining the region between 1 and 10 keV more
closely gives, with T varying in 1.0 keV increments, for the log of the TT
energy loss rate: -16.7, -5.1, -0.91, 1.36, 2.85, 3.94, 4.78, 5.48, 6.07, 6.57.
The total energy loss rate varies as: -15.96, -4.14, 0.37, 2.65, 3.77, 4.59,
5.25, 5.82, 6.31, 6.77. In short, at the RG, the TT energy loss rate, 1.4
erg/g$\cdot$s, is down by a factor 4.5 from its zero-density approximation
\cite{betal}, but the other processes bring the total rate up to 2.6
erg/g$\cdot$s, or within about a factor 2 of the zero density result.

Finally we turn to the dependence of the emission rates on the number n of
(large) extra dimensions.  We give, in Tables \ref{Tot3} and \ref{Tot4}, the
results for the total rates for n=3 and n=4.  One sees, in the low-density
limit, the $T^{7+n}$ behavior pointed out by Barger {\em et al.} \cite{betal}
for the $TT\to\mathcal{G}$ rate which dominates for $n=3,4$ in the same places
as in the $n=2$ case.

\section{Summary}

The approximations of zero density and purely transverse ($T+T\to\mathcal{G}$)
photon annihilation into gravitons of \cite{betal} must fail, for fixed
temperature, at high densities. We have computed a first estimate of finite
density corrections to two-plasmon production, for both transverse and
longitudinal plasmons of Kaluza-Klein excitations, as well as the decay process
$T\to L+{\cal G}$ as a function of plasma density and temperature over a wide
range of interest in both parameters. Our conclusion is that the zero-density,
pure transverse approximation is satisfactory for the sun, marginal for
supernovae, and fails by about a factor of 5 for red giants. It is interesting
to note that, while very little of the $\rho-T$ plane is occupied,
astrophysical systems appear to be preferentially located relatively close to
the boundaries (in $\rho$) at which the transverse photon approximation begins
to fail. Our calculation is approximate in that we omit most of the diagrams of
Fig.\,(\ref{diags}). However, we believe our results indicate that the full
calculation of all the diagrams is unlikely to modify our conclusions.

\begin{center}
\section*{\normalsize Acknowledgements}
\end{center}

One of us (V. T.) wishes to thank R. Mohapatra for helpful conversations. This
work was supported in part by the National Science Foundation under grant
PHY-9802439 and by the Department of Energy under Contract Nos.
DE-FG03-93ER40757 and DE-FG03-95ER40908.

\newpage
\begin{center}
\section*{References}
\end{center}

\newpage
\begin{center}
\section*{\small Tables}
\end{center}
\begin{table}[h]
\begin{center}
\fontfamily{cmr}\fontsize{10}{10}\selectfont
\begin{tabular}{|c|c|c|c|c|c|c|c|c|c|c|}\hline
$\rho\;\backslash\;T\sb$&1&$10^1$&$10^2$&$10^3$&$10^4$&$10^5$&$10^6$&$10^7$&$10^8$&$10^9$
\\ \hline
1 \sb   &- 47.34&-20.54 &-10.89&-1.88 &7.12&16.12&25.12&34.12&43.12&52.12\\ \hline
$10^1$\sb & -99.18&-23.76 &-10.97&-1.88 &7.12&16.12&25.12&34.12&43.12&52.12 \\ \hline
$10^2$\sb &-267.65&-38.33 &-11.54&-1.89 &7.12&16.12&25.12&34.12&43.12&52.12 \\ \hline
$10^3$\sb &       &-89.99 &-14.75&-1.97 &7.12&16.12&25.12&34.12&43.12&52.12 \\ \hline
$10^4$\sb &       &-255.96&-29.09&-2.53 &7.11&16.12&25.12&34.12&43.12&52.12 \\ \hline
$10^5$\sb &       &       &-77.48&-5.52 &7.03&16.12&25.12&34.12&43.12&52.12 \\ \hline
$10^6$\sb &       &       &-210.07&-16.77 &6.57 &16.11&25.12&34.12&43.12&52.12 \\ \hline
$10^7$\sb &       &       &     &-45.16 &4.96 &16.06&25.12&34.12&43.12&52.12 \\ \hline
$10^8$\sb &       &       &     &-106.83&0.36 &15.86&25.12&34.12&43.12&52.12 \\ \hline
$10^9\sb $&       &       &     &-239.38&-11.34&15.32&25.10&34.12&43.12&52.12 \\ \hline
$10^{10}$\sb&     &       &       &     &-38.58&13.77&25.03&34.12&43.12&52.12 \\ \hline
$10^{11}$\sb&     &       &       &     &-99.08& 9.25&24.84&34.12&43.12&52.12 \\ \hline
$10^{12}$\sb&     &       &       &     &-230.97&-2.40&24.32&34.10&43.12&52.12 \\ \hline
$10^{13}$\sb&     &       &       &       &     &-29.61&22.77&34.03&43.12&52.12\\ \hline
$10^{14}$\sb&     &       &       &       &     &-90.10&18.25&33.84&43.12&52.12\\ \hline
$10^{15}$\sb&     &       &       &       &     &-221.98&6.60&33.32&43.10&52.12\\ \hline
\end{tabular}
\end{center}
\vspace{-6pt}
\caption{\fontfamily{cmr}\fontsize{10}{10}\selectfont The entries are
$\log_{10}(\mathcal{Q})$
for the process (1) ($T+T\protect\to\mathcal{G}$). The rows are labeled by the density
$\rho$ in gm/cm$^3$, and the columns by the temperature $T$ in eV. A blank
entry indicates that $\mathcal{Q}<10^{-320}$\,erg/cm$^3$-s. The number of compact
dimensions is $n=2$.\label{TT}}
\end{table}
%\newpage
\begin{table}
\begin{center}
\fontfamily{cmr}\fontsize{10}{10}\selectfont
\begin{tabular}{|c|c|c|c|c|c|c|c|c|c|c|}\hline
$\rho\;\backslash\;T\sb$&1&$10^1$&$10^2$&$10^3$&$10^4$&$10^5$&$10^6$&$10^7$&$10^8$&$10^9$
\\ \hline
1 \sb       &-46.65&-20.73& -14.13& -8.11& -2.11&  9.39& 20.89& 30.05& 39.06& 48.07 \\ \hline
$10^1$\sb   &-98.74&-22.73& -12.70& -6.61& -0.61&  9.39& 20.89& 30.05& 39.06& 48.07 \\ \hline
$10^2$\sb  &-267.31&-37.64& -11.73& -5.13&  0.89&  9.39& 20.89& 30.05& 39.06& 48.07 \\ \hline
$10^3$\sb   &      &-89.55& -13.73& -3.70&  2.38&  9.40& 20.89& 30.05& 39.06& 48.07 \\ \hline
$10^4$\sb   &     &-255.62& -28.40& -2.74&  3.86&  9.87& 20.89& 30.05& 39.06& 48.07 \\ \hline
$10^5$\sb   &     &      & -77.05& -4.58&  5.27& 11.28& 20.89& 30.05& 39.06& 48.07 \\ \hline
$10^6$\sb   &     &      &-209.73&-16.14&  6.26& 12.75& 20.89& 30.05& 39.06& 48.07 \\ \hline
$10^7$\sb   &     &      &       &-44.73&  5.46& 14.04& 20.90& 30.05& 39.06& 48.07 \\ \hline
$10^8$\sb   &     &      &      &-106.47&  0.94& 14.99& 21.17& 30.05& 39.06& 48.07 \\ \hline
$10^9\sb $  &     &      &      &-239.05& -10.84&15.30& 22.25& 30.05& 39.06& 48.07 \\ \hline
$10^{10}$\sb&     &      &       &      & -38.17&14.21& 23.26& 30.05& 39.06& 48.07 \\ \hline
$10^{11}$\sb&     &      &       &      & -98.73& 9.81& 24.03& 30.28& 39.06& 48.07 \\ \hline
$10^{12}$\sb&     &      &       &      &-230.65&-1.90& 24.29& 31.29& 39.06& 48.07 \\ \hline
$10^{13}$\sb&     &      &       &      &      &-29.19& 23.20& 32.27& 39.06& 48.07 \\ \hline
$10^{14}$\sb&     &      &       &      &      &-89.74& 18.81& 33.03& 39.29& 48.07 \\ \hline
$10^{15}$\sb&     &      &       &      &      &-221.65& 7.10& 33.29& 40.31& 48.07 \\ \hline
\end{tabular}
\end{center}
\vspace{-6pt}
\caption{\fontfamily{cmr}\fontsize{10}{10}\selectfont The entries are
$\log_{10}(\mathcal{Q})$
for the process (2) ($T+L\protect\to\mathcal{G}$). The rows are labeled by the density
$\rho$ in gm/cm$^3$, and the columns by the temperature $T$ in eV. The number of compact
dimensions is $n=2$.\label{TL}}
\end{table}
\newpage
\begin{table}[h]
\begin{center}
\fontfamily{cmr}\fontsize{10}{10}\selectfont
\begin{tabular}{|c|c|c|c|c|c|c|c|c|c|c|}\hline
$\rho\;\backslash\;T\sb$&1&$10^1$&$10^2$&$10^3$&$10^4$&$10^5$&$10^6$&$10^7$&$10^8$&$10^9$
\\ \hline
1\sb    &\hb     &-20.43&-16.94&-15.08&-14.01&  1.16& 15.49& 24.75& 33.78& 42.81 \\ \hline
$10^1$\sb   &    &-20.91&-13.70&-11.58& -9.96&  1.16& 15.49& 24.75& 33.78& 42.81 \\ \hline
$10^2$\sb   &    &      &-11.44& -8.09& -6.41&  1.16& 15.49& 24.75& 33.78& 42.81 \\ \hline
$10^3$\sb   &    &      &-11.92& -4.84& -2.93&  1.18& 15.49& 24.75& 33.78& 42.81 \\ \hline
$10^4$\sb   &    &      &      & -2.58&  0.46&  2.29& 15.49& 24.75& 33.78& 42.81 \\ \hline
$10^5$\sb   &    &      &      & -3.15&  3.31&  5.47& 15.49& 24.75& 33.78& 42.81 \\ \hline
$10^6$\sb   &    &      &      &      &  5.33&  8.82& 15.49& 24.75& 33.78& 42.81 \\ \hline
$10^7$\sb   &    &      &      &      &  4.84& 10.97& 15.49& 24.75& 33.78& 42.81 \\ \hline
$10^8$\sb   &    &      &      &      &      & 12.64& 16.10& 24.75& 33.78& 42.81 \\ \hline
$10^9\sb $  &    &      &      &      &      & 13.26& 18.09& 24.75& 33.78& 42.81 \\ \hline
$10^{10}$\sb&    &      &      &      &      &      & 19.99& 24.74& 33.78& 42.81 \\ \hline
$10^{11}$\sb&    &      &      &      &      &      & 21.36& 25.18& 33.78& 42.81 \\ \hline
$10^{12}$\sb&    &      &      &      &      &      & 19.86& 27.14& 33.78& 42.81 \\ \hline
$10^{13}$\sb&    &      &      &      &      &      &      & 28.88& 33.77& 42.81 \\ \hline
$10^{14}$\sb&    &      &      &      &      &      &      & 29.94& 34.23& 42.81 \\ \hline
$10^{15}$\sb&    &      &      &      &      &      &      & 27.24& 36.13& 42.81 \\ \hline
\end{tabular}
\end{center}
\vspace{-6pt}
\caption{\fontfamily{cmr}\fontsize{10}{10}\selectfont The entries are
$\log_{10}(\mathcal{Q})$
for the process (4) ($T\protect\to L+\mathcal{G}$). The rows are labeled by the density
$\rho$ in gm/cm$^3$, and the columns by the temperature $T$ in eV. The number of compact
dimensions is $n=2$.\label{TtoL}}
\end{table}
\newpage
\begin{table}[h]
\begin{center}
\fontfamily{cmr}\fontsize{10}{10}\selectfont
\begin{tabular}{|c|c|c|c|c|c|c|c|c|c|c|}\hline
$\rho\;\backslash\;T\sb$&1&$10^1$&$10^2$&$10^3$&$10^4$&$10^5$&$10^6$&$10^7$&$10^8$&$10^9$
\\ \hline
1 \sb       &  -47.16&-21.95&-18.57&-16.46&-14.47&  0.37& 15.21& 24.59& 33.63& 42.68 \\ \hline
$10^1$\sb   &  -99.50&-22.88&-15.35&-12.99&-10.97&  0.37& 15.21& 24.59& 33.63& 42.68 \\ \hline
$10^2$\sb   & -268.17&-38.15&-12.95& -9.57& -7.48&  0.37& 15.21& 24.59& 33.63& 42.68 \\ \hline
$10^3$\sb   &        &-90.32&-13.88& -6.35& -4.01&  0.39& 15.21& 24.59& 33.63& 42.68 \\ \hline
$10^4$\sb   &       &-256.47&-28.92& -3.96& -0.59&  1.49& 15.21& 24.59& 33.63& 42.68 \\ \hline
$10^5$\sb   &        &      &-77.82& -4.80&  2.58&  4.76& 15.21& 24.59& 33.63& 42.68 \\ \hline
$10^6$\sb   &        &     &-210.59&-16.71&  4.96&  8.16& 15.21& 24.59& 33.63& 42.68 \\ \hline
$10^7$\sb   &        &      &      &-45.50&  4.83& 11.09& 15.22& 24.59& 33.63& 42.68 \\ \hline
$10^8$\sb   &        &      &     &-107.32&  0.32& 13.25& 15.85& 24.59& 33.63& 42.68 \\ \hline
$10^9\sb $  &        &      &     &-239.92&-11.54& 14.26& 18.34& 24.59& 33.63& 42.68 \\ \hline
$10^{10}$\sb&        &      &      &      &-38.96& 13.49& 20.62& 24.59& 33.63& 42.68 \\ \hline
$10^{11}$\sb&        &      &      &      &-99.58&  9.18& 22.35& 25.12& 33.63& 42.68 \\ \hline
$10^{12}$\sb&        &      &      &     &-231.52& -2.60& 23.25& 27.45& 33.63& 42.68 \\ \hline
$10^{13}$\sb&        &      &      &      &      &-29.98& 22.49& 29.65& 33.63& 42.68 \\ \hline
$10^{14}$\sb&        &      &      &      &      &-90.59& 18.18& 31.36& 34.17& 42.68 \\ \hline
$10^{15}$\sb&        &      &      &      &     &-222.53&  6.40& 32.25& 36.49& 42.68 \\ \hline
\end{tabular}
\end{center}
\vspace{-6pt}
\caption{\fontfamily{cmr}\fontsize{10}{10}\selectfont The entries are
$\log_{10}(\mathcal{Q})$
for the process (3) ($L+L\protect\to\mathcal{G}$). The rows are labeled by the density
$\rho$ in gm/cm$^3$, and the columns by the temperature $T$ in eV. The number of compact
dimensions is $n=2$.\label{LL}}
\end{table}
\newpage
\begin{table}[h]
\begin{center}
\fontfamily{cmr}\fontsize{10}{10}\selectfont
\begin{tabular}{|c|c|c|c|c|c|c|c|c|c|c|}\hline
$\rho\;\backslash\;T\sb$&1&$10^1$&$10^2$&$10^3$&$10^4$&$10^5$&$10^6$&$10^7$&$10^8$&$10^9$
\\ \hline
1 \sb       &  -46.47&-20.07&-10.89& -1.88&  7.12& 16.12& 25.12& 34.12& 43.12& 52.12 \\ \hline
$10^1$\sb   &  -98.55&-20.90&-10.96& -1.88&  7.12& 16.12& 25.12& 34.12& 43.12& 52.12 \\ \hline
$10^2$\sb   & -267.11&-37.46&-11.07& -1.89&  7.12& 16.12& 25.12& 34.12& 43.12& 52.12 \\ \hline
$10^3$\sb   &        &-89.37&-11.91& -1.96&  7.12& 16.12& 25.12& 34.12& 43.12& 52.12 \\ \hline
$10^4$\sb   &       & -255.41&-28.22&-2.12&  7.11& 16.12& 25.12& 34.12& 43.12& 52.12 \\ \hline
$10^5$\sb   &        &      &-76.86& -3.12&  7.04& 16.12& 25.12& 34.12& 43.12& 52.12 \\ \hline
$10^6$\sb   &        &     &-209.53&-15.96&  6.77& 16.12& 25.12& 34.12& 43.12& 52.12 \\ \hline
$10^7$\sb   &        &      &      &-44.54&  5.72& 16.06& 25.12& 34.12& 43.12& 52.12 \\ \hline
$10^8$\sb   &        &      &     &-106.27&  1.11& 15.91& 25.12& 34.12& 43.12& 52.12 \\ \hline
$10^9\sb $  &        &      &     &-238.84&-10.66& 15.63& 25.10& 34.12& 43.12& 52.12 \\ \hline
$10^{10}$\sb&        &      &      &      &-37.98& 14.40& 25.04& 34.12& 43.12& 52.12 \\ \hline
$10^{11}$\sb&        &      &      &      &-98.53&  9.99& 24.90& 34.12& 43.12& 52.12 \\ \hline
$10^{12}$\sb&        &      &      &     &-230.44& -1.72& 24.63& 34.10& 43.12& 52.12 \\ \hline
$10^{13}$\sb&        &      &      &      &      &-25.04& 23.40& 34.04& 43.12& 52.12 \\ \hline
$10^{14}$\sb&        &      &      &      &      &-89.54& 18.99& 33.90& 43.12& 52.12 \\ \hline
$10^{15}$\sb&        &      &      &      &     &-221.45&  7.28& 33.63& 43.10& 52.12 \\ \hline
\end{tabular}
\end{center}
\vspace{-6pt}
\caption{\fontfamily{cmr}\fontsize{10}{10}\selectfont The entries are
$\log_{10}(\mathcal{Q})$
for the sum of processes (1)-(4). The rows are labeled by the density
$\rho$ in gm/cm$^3$, and the columns by the temperature $T$ in eV. The number of compact
dimensions is $n=2$.\label{Tot2}}
\end{table}
\newpage
\begin{table}[h]
\begin{center}
\fontfamily{cmr}\fontsize{10}{10}\selectfont
\begin{tabular}{|c|c|c|c|c|c|c|c|c|c|c|}\hline
$\rho\;\backslash\;T\sb$&1&$10^1$&$10^2$&$10^3$&$10^4$&$10^5$&$10^6$&$10^7$&$10^8$&$10^9$
\\ \hline
1 \sb       &2&  3&  1&  1&  1&  1&  1&  1&  1&  1  \\ \hline
$10^1$\sb   &2&  3&  1&  1&  1&  1&  1&  1&  1&  1  \\ \hline
$10^2$\sb   &2&  2&  3&  1&  1&  1&  1&  1&  1&  1  \\ \hline
$10^3$\sb   &0&  2&  3&  1&  1&  1&  1&  1&  1&  1  \\ \hline
$10^4$\sb   &0&  2&  2&  1&  1&  1&  1&  1&  1&  1  \\ \hline
$10^5$\sb   &0&  0&  2&  3&  1&  1&  1&  1&  1&  1  \\ \hline
$10^6$\sb   &0&  0&  2&  2&  1&  1&  1&  1&  1&  1  \\ \hline
$10^7$\sb   &0&  0&  0&  2&  2&  1&  1&  1&  1&  1  \\ \hline
$10^8$\sb   &0&  0&  0&  2&  2&  1&  1&  1&  1&  1  \\ \hline
$10^9\sb $  &0&  0&  0&  2&  2&  1&  1&  1&  1&  1  \\ \hline
$10^{10}$\sb&0&  0&  0&  0&  2&  2&  1&  1&  1&  1  \\ \hline
$10^{11}$\sb&0&  0&  0&  0&  2&  2&  1&  1&  1&  1  \\ \hline
$10^{12}$\sb&0&  0&  0&  0&  2&  2&  1&  1&  1&  1  \\ \hline
$10^{13}$\sb&0&  0&  0&  0&  0&  3&  2&  1&  1&  1  \\ \hline
$10^{14}$\sb&0&  0&  0&  0&  0&  2&  2&  1&  1&  1  \\ \hline
$10^{15}$\sb&0&  0&  0&  0&  0&  2&  2&  1&  1&  1  \\ \hline
\end{tabular}
\end{center}
\vspace{-6pt}
\caption{\fontfamily{cmr}\fontsize{10}{10}\selectfont The entries denote the
the process which makes the dominant contribution to $\mathcal{Q}$. A zero
indicates that all of the processes are negligible
($\mathcal{Q}<10^{-320}$\,erg/cm$^3$-s). \label{dom}}
\end{table}
\newpage
\begin{table}[h]
\begin{center}
\fontfamily{cmr}\fontsize{10}{10}\selectfont
\begin{tabular}{|c|c|c|c|c|c|c|c|c|c|c|}\hline
$\rho\;\backslash\;T\sb$&1&$10^1$&$10^2$&$10^3$&$10^4$&$10^5$&$10^6$&$10^7$&$10^8$&$10^9$
\\ \hline
1 \sb       &0.66&   0.43& 1.00&   1.00&   1.00&   1.00&   1.00&   1.00&   1.00&1.00  \\ \hline
$10^1$\sb   &0.65&   0.97& 0.98&   1.00&   1.00&   1.00&   1.00&   1.00&   1.00&1.00  \\ \hline
$10^2$\sb   &0.63&   0.66& 0.43&   1.00&   1.00&   1.00&   1.00&   1.00&   1.00&1.00  \\ \hline
$10^3$\sb   &0.00&   0.65& 0.97&   0.98&   1.00&   1.00&   1.00&   1.00&   1.00&1.00  \\ \hline
$10^4$\sb   &0.00&   0.63& 0.66&   0.39&   1.00&   1.00&   1.00&   1.00&   1.00&1.00  \\ \hline
$10^5$\sb   &0.00&   0.00& 0.65&   0.94&   0.98&   1.00&   1.00&   1.00&   1.00&1.00  \\ \hline
$10^6$\sb   &0.00&   0.00& 0.63&   0.67&   0.64&   1.00&   1.00&   1.00&   1.00&1.00  \\ \hline
$10^7$\sb   &0.00&   0.00& 0.00&   0.65&   0.56&   0.99&   1.00&   1.00&   1.00&1.00  \\ \hline
$10^8$\sb   &0.00&   0.00& 0.00&   0.63&   0.66&   0.88&   1.00&   1.00&   1.00&1.00  \\ \hline
$10^9\sb $  &0.00&   0.00& 0.00&   0.62&   0.66&   0.49&   1.00&   1.00&   1.00&1.00  \\ \hline
$10^{10}$\sb&0.00&   0.00& 0.00&   0.00&   0.65&   0.64&   0.98&   1.00&   1.00&1.00  \\ \hline
$10^{11}$\sb&0.00&   0.00& 0.00&   0.00&   0.63&   0.66&   0.86&   1.00&   1.00&1.00  \\ \hline
$10^{12}$\sb&0.00&   0.00& 0.00&   0.00&   0.62&   0.66&   0.49&   1.00&   1.00&1.00  \\ \hline
$10^{13}$\sb&0.00&   0.00& 0.00&   0.00&   0.00&   1.00&   0.64&   0.98&   1.00&1.00  \\ \hline
$10^{14}$\sb&0.00&   0.00& 0.00&   0.00&   0.00&   0.63&   0.66&   0.86&   1.00&1.00  \\ \hline
$10^{15}$\sb&0.00&   0.00& 0.00&   0.00&   0.00&   0.62&   0.66&   0.49&   1.00&1.00  \\ \hline
\end{tabular}
\end{center}
\vspace{-6pt}
\caption{\fontfamily{cmr}\fontsize{10}{10}\selectfont The entries are the
fraction of the contribution to $\mathcal{Q}$ due to the dominant process of
Table \ref{dom}.\label{frac}}
\end{table}
\newpage
\begin{table}[h]
\begin{center}
\fontfamily{cmr}\fontsize{10}{10}\selectfont
\begin{tabular}{|c|c|c|c|c|c|c|c|c|c|c|}\hline
$\rho\;\backslash\;T\sb$&1&$10^1$&$10^2$&$10^3$&$10^4$&$10^5$&$10^6$&$10^7$&$10^8$&$10^9$
\\ \hline
1 \sb       &  -56.71& -30.35&  -20.05& -10.04&  -0.04&    9.96& 19.96& 29.95& 39.95& 49.95   \\ \hline
$10^1$\sb   & -108.29& -31.47&  -20.11& -10.04&  -0.04&    9.96& 19.96& 29.95& 39.95& 49.95   \\ \hline
$10^2$\sb   & -276.35& -46.70&  -20.36& -10.05&  -0.04&    9.96& 19.96& 29.95& 39.95& 49.95   \\ \hline
$10^3$\sb   &        & -98.11&  -21.48& -10.10&  -0.04&    9.96& 19.96& 29.95& 39.95& 49.95   \\ \hline
$10^4$\sb   &        &-263.66&  -36.47& -10.36&  -0.05&    9.96& 19.96& 29.95& 39.95& 49.95   \\ \hline
$10^5$\sb   &        &       &  -84.62& -11.64&  -0.10&    9.96& 19.96& 29.95& 39.95& 49.95   \\ \hline
$10^6$\sb   &        &       & -216.84& -23.28&  -0.35&    9.95& 19.96& 29.95& 39.95& 49.95   \\ \hline
$10^7$\sb   &        &       &        & -51.47&  -1.22&    9.91& 19.96& 29.95& 39.95& 49.95   \\ \hline
$10^8$\sb   &        &       &        &-112.85&  -5.45&    9.78& 19.95& 29.95& 39.95& 49.95   \\ \hline
$10^9\sb $  &        &       &        &-245.09& -16.90&    9.56& 19.94& 29.95& 39.95& 49.95   \\ \hline
$10^{10}$\sb&        &       &        &       & -43.88&    8.54& 19.89& 29.95& 39.95& 49.95   \\ \hline
$10^{11}$\sb&        &       &        &       &-104.10&    4.43& 19.77& 29.95& 39.95& 49.95   \\ \hline
$10^{12}$\sb&        &       &        &       &-235.68&   -6.96& 19.55& 29.94& 39.95& 49.95   \\ \hline
$10^{13}$\sb&        &       &        &       &       &  -33.90& 18.54& 29.89& 39.95& 49.95   \\ \hline
$10^{14}$\sb&        &       &        &       &       &  -94.12& 14.43& 29.77& 39.95& 49.95   \\ \hline
$10^{15}$\sb&        &       &        &       &       & -225.69&  3.04& 29.55& 39.94& 49.95   \\ \hline
\end{tabular}
\end{center}
\vspace{-6pt}
\caption{\fontfamily{cmr}\fontsize{10}{10}\selectfont The entries are $\log_{10}(\mathcal{Q})$
for the sum of processes (1)-(4) in the case of $n=3$ compact dimensions. The rows are labeled
by the density $\rho$ in gm/cm$^3$, and the columns by the temperature $T$ in eV. \label{Tot3}}
\end{table}
\newpage
\begin{table}[h]
\begin{center}
\fontfamily{cmr}\fontsize{10}{10}\selectfont
\begin{tabular}{|c|c|c|c|c|c|c|c|c|c|c|}\hline
$\rho\;\backslash\;T\sb$&1&$10^1$&$10^2$&$10^3$&$10^4$&$10^5$&$10^6$&$10^7$&$10^8$&$10^9$
\\ \hline
1 \sb       & -66.96& -40.44& -29.15& -18.15&  -7.15&   3.85&14.85&25.85&36.85&47.85   \\ \hline
$10^1$\sb   &-118.03& -41.77& -29.20& -18.15&  -7.15&   3.85&14.85&25.85&36.85&47.85   \\ \hline
$10^2$\sb   &-285.59& -55.95& -29.44& -18.15&  -7.15&   3.85&14.85&25.85&36.85&47.85   \\ \hline
$10^3$\sb   &       &-106.85& -30.77& -18.20&  -7.15&   3.85&14.85&25.85&36.85&47.85   \\ \hline
$10^4$\sb   &       &-271.91& -44.72& -18.44&  -7.15&   3.85&14.85&25.85&36.85&47.85   \\ \hline
$10^5$\sb   &       &       & -92.38& -19.79&  -7.19&   3.85&14.85&25.85&36.85&47.85   \\ \hline
$10^6$\sb   &       &       &-224.16& -30.59&  -7.41&   3.85&14.85&25.85&36.85&47.85   \\ \hline
$10^7$\sb   &       &       &       & -58.40&  -8.11&   3.82&14.85&25.85&36.85&47.85   \\ \hline
$10^8$\sb   &       &       &       &-119.43& -12.02&   3.70&14.85&25.85&36.85&47.85   \\ \hline
$10^9\sb $  &       &       &       &-251.33& -23.14&   3.51&14.84&25.85&36.85&47.85   \\ \hline
$10^{10}$\sb&       &       &       &       & -49.79&   2.69&14.80&25.85&36.85&47.85   \\ \hline
$10^{11}$\sb&       &       &       &       &-109.68&  -1.13&14.69&25.85&36.85&47.85   \\ \hline
$10^{12}$\sb&       &       &       &       &-240.93& -12.19&14.51&25.83&36.85&47.85   \\ \hline
$10^{13}$\sb&       &       &       &       &       & -38.82&13.69&25.79&36.85&47.85   \\ \hline
$10^{14}$\sb&       &       &       &       &       & -98.69& 9.87&25.69&36.85&47.85   \\ \hline
$10^{15}$\sb&       &       &       &       &       &-229.93&-1.19&25.51&36.83&47.85   \\ \hline
\end{tabular}
\end{center}
\vspace{-6pt}
\caption{\fontfamily{cmr}\fontsize{10}{10}\selectfont The entries are $\log_{10}(\mathcal{Q})$
for the sum of processes (1)-(4) in the case of $n=4$ compact dimensions. The rows are labeled
by the density $\rho$ in gm/cm$^3$, and the columns by the temperature $T$ in eV. \label{Tot4}}
\end{table}
\newpage
\begin{table}[h]
\begin{center}
\fontfamily{cmr}\fontsize{9}{10}\selectfont
\begin{tabular}{|c|c|c|c|c|c|c|c|c|c|c|}\hline
$\rho\;\backslash\;T\sb$&1&$10^1$&$10^2$&$10^3$&$10^4$&$10^5$&$10^6$&$10^7$&$10^8$&$10^9$
\\ \hline
1 \sb       &0.51  &0.51  &0.51  & 0.50  & 0.42  &1.8\t{-5} &1.4\t{-8} &1.4\t{-10} &1.4\t{-12}& 2.8\t{-14}  \\ \hline
$10^1$\sb   &0.51  &0.51  &0.51  & 0.51  & 0.44  &1.8\t{-4} &1.4\t{-7} &1.4\t{-9}  &1.4\t{-11}& 1.4\t{-13}  \\ \hline
$10^2$\sb   &0.51  &0.51  &0.51  & 0.51  & 0.46  &1.8\t{-3} &1.4\t{-6} &1.4\t{-8}  &1.4\t{-10}& 1.4\t{-12}  \\ \hline
$10^3$\sb   &0.51  &0.51  &0.51  & 0.51  & 0.49  &0.02      &1.4\t{-5} &1.4\t{-7}  &1.4\t{-9} & 1.4\t{-11}  \\ \hline
$10^4$\sb   &0.52  &0.52  &0.52  & 0.52  & 0.51  &0.14      &1.4\t{-4} &1.4\t{-6}  &1.4\t{-8} & 1.4\t{-10}  \\ \hline
$10^5$\sb   &0.56  &0.56  &0.56  & 0.56  & 0.56  &0.37      &1.4\t{-3} &1.4\t{-5}  &1.4\t{-7} & 1.4\t{-9}   \\ \hline
$10^6$\sb   &0.72  &0.72  &0.72  & 0.72  & 0.72  &0.65      &0.014     &1.4\t{-4}  &1.4\t{-6} & 1.4\t{-8}   \\ \hline
$10^7$\sb   &1.20  &1.20  &1.20  & 1.20  & 1.20  &1.20      &0.14      &1.4\t{-3}  &1.4\t{-5} & 1.4\t{-7}   \\ \hline
$10^8$\sb   &2.40  &2.40  &2.40  & 2.40  & 2.40  &2.40      &1.20      &0.014      &1.4\t{-4} & 1.4\t{-6}   \\ \hline
$10^9\sb $  &5.20  &5.20  &5.20  & 5.20  & 5.20  &5.20      &4.50      &0.14       &1.4\t{-3} & 1.4\t{-5}   \\ \hline
$10^{10}$\sb&11.0  &11.0  &11.0  & 11.0  & 11.0  &11.0      &11.0      &1.40       &0.014     & 1.4\t{-4}   \\ \hline
$10^{11}$\sb&24.0  &24.0  &24.0  & 24.0  & 24.0  &24.0      &24.0      &12.0       &0.14      & 1.4\t{-3}   \\ \hline
$10^{12}$\sb&51.0  &51.0  &51.0  & 51.0  & 51.0  &51.0      &51.0      &45.0       &1.40      & 0.014       \\ \hline
$10^{13}$\sb&110.0 &110.0 &110.0 & 110.0 & 110.0 &110.0     &110.0     &110.0      &14.0      & 0.14        \\ \hline
$10^{14}$\sb&240.0 &240.0 &240.0 & 240.0 & 240.0 &240.0     &240.0     &240.0      &120.0     & 1.40        \\ \hline
$10^{15}$\sb&510.0 &510.0 &510.0 & 510.0 & 510.0 &510.0     &510.0     &510.0      &450.0     & 14.0        \\ \hline
\end{tabular}
\end{center}
\vspace{-6pt}
\caption{\fontfamily{cmr}\fontsize{10}{10}\selectfont The entries are the
chemical potential $\mu(\rho,T)$. The rows are labeled
by the density $\rho$ in gm/cm$^3$, and the columns by the temperature $T$ in eV. \label{mu}}
\end{table}
\newpage
\begin{table}[h]
\begin{center}
\fontfamily{cmr}\fontsize{9}{10}\selectfont
\begin{tabular}{|c|c|c|c|c|c|c|c|c|c|c|}\hline
$\rho\;\backslash\;T\sb$&1&$10^1$&$10^2$&$10^3$&$10^4$&$10^5$&$10^6$&$10^7$&$10^8$&$10^9$
\\ \hline
1 \sb       &2.6\t{-5}&2.2\t{-5}& -2.3\t{-4}& -5.8\t{-3}& -0.09 &  -0.51 &  -0.51 &  -0.51 &  -0.51 &  -0.51 \\ \hline
$10^1$\sb   &1.2\t{-4}&1.2\t{-4}&  3.6\t{-5}& -3.5\t{-3}& -0.07 &  -0.51 &  -0.51 &  -0.51 &  -0.51 &  -0.51 \\ \hline
$10^2$\sb   &5.6\t{-4}&5.6\t{-4}&  5.4\t{-4}& -1.1\t{-3}& -0.05 &  -0.51 &  -0.51 &  -0.51 &  -0.51 &  -0.51 \\ \hline
$10^3$\sb   &2.6\t{-3}&2.6\t{-3}&  2.6\t{-3}&  2.2\t{-3}& -0.02 &  -0.49 &  -0.51 &  -0.51 &  -0.51 &  -0.51 \\ \hline
$10^4$\sb   &0.01     &0.01     &  0.01     &  0.01     &  0.00 &  -0.37 &  -0.51 &  -0.51 &  -0.51 &  -0.51 \\ \hline
$10^5$\sb   &0.05     &0.05     &  0.05     &  0.05     &  0.05 &  -0.14 &  -0.51 &  -0.51 &  -0.51 &  -0.51 \\ \hline
$10^6$\sb   &0.21     &0.21     &  0.21     &  0.21     &  0.21 &   0.14 &  -0.50 &  -0.51 &  -0.51 &  -0.51 \\ \hline
$10^7$\sb   &0.71     &0.71     &  0.71     &  0.71     &  0.71 &   0.68 &  -0.37 &  -0.51 &  -0.51 &  -0.51 \\ \hline
$10^8$\sb   &1.90     &1.90     &  1.90     &  1.90     &  1.90 &   1.90 &   0.72 &  -0.50 &  -0.51 &  -0.51 \\ \hline
$10^9\sb $  &4.60     &4.60     &  4.60     &  4.60     &  4.60 &   4.60 &   4.00 &  -0.37 &  -0.51 &  -0.51 \\ \hline
$10^{10}$\sb&11.0     &11.0     &  11.0     &  11.0     &  11.0 &   11.0 &   10.0 &   0.86 &  -0.50 &  -0.51 \\ \hline
$10^{11}$\sb&23.0     &23.0     &  23.0     &  23.0     &  23.0 &   23.0 &   23.0 &   11.0 &  -0.37 &  -0.51 \\ \hline
$10^{12}$\sb&51.0     &51.0     &  51.0     &  51.0     &  51.0 &   51.0 &   51.0 &   44.0 &   0.87 &  -0.50 \\ \hline
$10^{13}$\sb&110.0    &110.0    &  110.0    &  110.0    &  110.0&   110.0&   110.0&   110.0&   13.0 &  -0.37 \\ \hline
$10^{14}$\sb&240.0    &240.0    &  240.0    &  240.0    &  240.0&   240.0&   240.0&   240.0&   120.0&   0.87 \\ \hline
$10^{15}$\sb&510.0    &510.0    &  510.0    &  510.0    &  510.0&   510.0&   510.0&   510.0&   450.0&   13.0 \\ \hline
\end{tabular}
\end{center}
\vspace{-6pt}
\caption{\fontfamily{cmr}\fontsize{10}{10}\selectfont The entries are $\tilde{\mu}(\rho,T)$
the difference between the chemical potential $\mu(\rho,T)$ and $m_e$. The rows are labeled
by the density $\rho$ in gm/cm$^3$, and the columns by the temperature $T$ in eV.
\label{mutilde}}
\end{table}


\begin{thebibliography}{99}
\bibitem{betal} V. Barger, T. Han, C. Kao and R.-J. Zhang, Phys. Lett. {\bf B461},
34 (1999); hep-ph/9905474 (1999).
\bibitem{add} N. Arkani-Hamed, S. Dimopoulos and G. Dvali, Phys. Lett. B {\bf
429}, 263 (1998); I. Antoniadis, N. Arkani-Hamed, S. Dimopoulos and G. Dvali
{\it ibid.}, {\bf 436}, 506 (1998).
\bibitem{weinberg} S. Weinberg, {\it Gravitation and Cosmology}, John Wiley and
Sons, New York, 1972.
\bibitem{grw} G. F. Guidice, R. Ratazzi and J. D. Wells, Nucl. Phys. B {\bf
544}, 3 (1999).
\bibitem{hlz} T. Han, J. D. Lykken and R.-J. Zhang, Phys. Rev. D {\bf 59},
105006 (1999).
\bibitem{bs} E. Bratten and D. Segel, Phys. Rev. D {\bf 48}, 1478 (1993).
\bibitem{raffelt} Georg G. Raffelt, {\em Stars as Laboratories for Fundamental
Physics}, The University of Chicago Press, 1996.
\end{thebibliography}
\end{document}